%% file: main.tex
\pdfoutput=1

\documentclass[runningheads]{llncs}
\usepackage{ifpdf}
\usepackage{graphicx}
\usepackage{tikz}
\usepackage{amsmath}
\usepackage{cite}
\usepackage{amsmath,amssymb,amsfonts}
\usepackage{algorithm}
\usepackage{algpseudocode}
\usepackage{textcomp}
\usepackage{xcolor}
\usepackage{tabularx}
\usepackage{verbatim}
\usepackage{multirow}
\PassOptionsToPackage{hyphens}{url}\usepackage{hyperref}
\usepackage{balance}
\usepackage[toc]{appendix}
\usepackage{caption}
\usepackage{subcaption}
\usepackage{longtable}
\usepackage{balance}
\usepackage{epstopdf}
\usepackage[T1]{fontenc}
\usepackage[utf8x]{inputenc}

\newcommand{\ignore}[1]{}
\newcommand{\systemName}{Shadow-Catcher}
\begin{document}
\title{Shadow-Catcher: Looking Into Shadows to Detect Ghost Objects in Autonomous Vehicle 3D Sensing}
\titlerunning{Shadow-Catcher: Detecting Ghost Objects in AV 3D Sensing with Shadows}
%

\author{Zhongyuan Hau \and
Soteris Demetriou \and
Luis Muñoz-González \and
Emil C. Lupu
}
\authorrunning{Z. Hau et al.}
%
\institute{{Imperial College London, United Kingdom} \\
\email{\{zy.hau17, s.demetriou, l.munoz, e.c.lupu\}@imperial.ac.uk}}

\maketitle              

\begin{abstract}
\input{sections/0_abstract}

\end{abstract}

\section{Introduction}
\input{sections/1_introduction}

\section{Background and Related Work}
\label{sec:related}
\input{sections/2_background}

\section{Threat Model}
\label{sec:threat_model}
\input{sections/2_1_threat_model}

\section{3D Shadows as a Physical Invariant}
\label{sec:obj_shadow}
\input{sections/3_object_to_shadows}

\section{Shadow-Catcher Design}
\label{sec:system}
\input{sections/4_ghostbuster}

\section{Evaluation}
\label{sec:results_evaluation}
\input{sections/5_results_evaluation}

\section{Conclusion}
\label{sec:conclusion}
\input{sections/8_conclusion}

\balance
\bibliographystyle{plain}
\bibliography{biblio}

\appendix
\section*{Appendices}
\input{appendix/appendix}

\end{document}

%% file: sections/0_abstract.tex
LiDAR-driven 3D sensing allows new generations of vehicles to achieve advanced levels of situation awareness. However, recent works have demonstrated that physical adversaries can spoof LiDAR return signals and deceive 3D object detectors to erroneously detect ``ghost" objects. Existing defenses are either impractical or focus only on vehicles. Unfortunately, it is easier to spoof smaller objects such as pedestrians and cyclists, but harder to defend against and can have worse safety implications. To address this gap, we introduce \systemName{}, a set of new techniques embodied in an end-to-end prototype to detect both large and small ghost object attacks on 3D detectors. We characterize a new semantically meaningful physical invariant (3D shadows) which \systemName{} leverages for validating objects. Our evaluation on the KITTI dataset shows that \systemName{} consistently achieves more than 94\% accuracy in identifying anomalous shadows for vehicles, pedestrians, and cyclists, while it remains robust to a novel class of strong ``invalidation'' attacks targeting the defense system. \systemName{} can achieve real-time detection, requiring only between 0.003s–0.021s on average to process an object in a 3D point cloud on commodity hardware and achieves a 2.17x speedup compared to prior work.

%% file: sections/1_introduction.tex
High-precision depth sensors are increasingly being used for mapping the environment in various application domains, such as robotics~\cite{staff_2019}, security surveillance~\cite{gips_2020} and augmented reality applications~\cite{porter_2020}. LiDARs (derived from light detection and ranging) are popular such depth sensors. They are pervasively deployed~\cite{coldewey_2018, bbc_news_2019} in autonomous vehicles (referred to as AVs henceforth) where a new class of Deep Neural Network (DNN) 3D classifiers leverage their measurements (processed in batches called 3D point clouds) to detect objects -- a necessary task for downstream safety-critical driving decision-making \cite{qi2017pointnet++, DBLP:journals/corr/abs-1902-06326, Shi2020PointGNNGN, shi2020pv}.

Recent studies have shown that it is possible to attack LiDAR-based perception systems by spoofing LiDAR return signals~\cite{petit2015remote, shin2017illusion, cao2019adversarial}. To defend against model-level LiDAR spoofing attacks, prior works suggested using sensor fusion~\cite{cao2019adversarial}, view fusion (SVF)~\cite{255240}, or leveraging 3D-point statistical anomaly detection based on physical invariants such as object occlusions and free space (CARLO)~\cite{255240}. Unfortunately, sensor fusion approaches~\cite{ivanov2014attack, yang2018sensor} rely on the assumption that a majority of the sensors are not under attack. SVF makes fewer assumptions but requires expensive retraining of the classifiers and has reduced classification accuracy, which is more dangerous than failing to detect ghost objects. CARLO is a backward compatible method, agnostic to the adversary and achieves good accuracy in detecting spoofed vehicles with an acceptable performance overhead. However, CARLO depends on the size of the object's bounding box and on the fact that genuine vehicles exhibit a high points' density. This approach does not work for smaller objects such as pedestrians and cyclists. Lastly, there is no approach to date which uses semantically meaningful information which can be crucial in reasoning and explaining system decisions.

\begin{figure}[t]
    \centering
    \begin{minipage}{0.45\textwidth}
        \centering
        \includegraphics[width=0.9\textwidth,trim={0 9cm 0 5cm},clip]{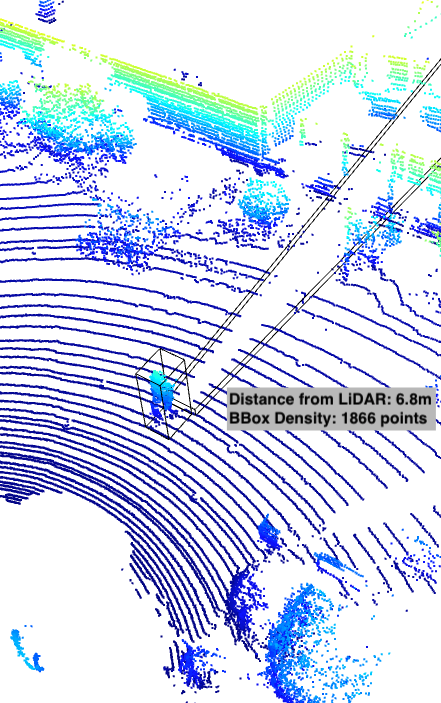} 
        \caption{Genuine object's \textit{3D Shadow}}
         \label{fig:real_ped}
    \end{minipage}\hfill
    \begin{minipage}{0.45\textwidth}
        \centering
        \includegraphics[width=0.9\textwidth,trim={0 6cm 0 4cm},clip]{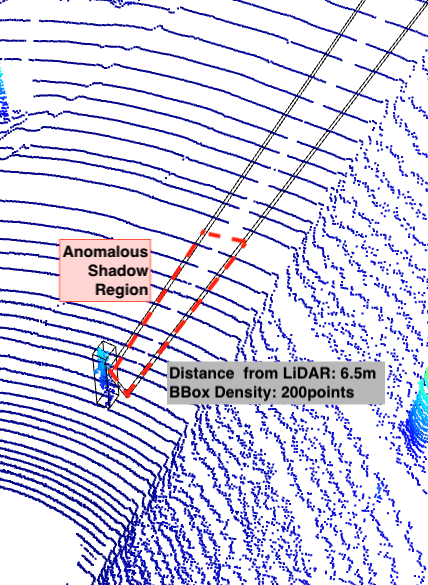} 
        \caption{Ghost object's \textit{3D Shadow}}
        \label{fig:fake_ped}
    \end{minipage}
    \vspace{-6mm}
\end{figure}

In this work, we introduce a new approach to detect model-level LiDAR spoofing attacks on both large and small objects that assumes neither the presence nor the cooperation of other sensors. Our mechanism is agnostic to the classification model targeted: any detected object, either genuine or fake (\emph{ghost}), will be subjected to verification. We observe that real 3D objects are closely followed by 3D shadows, which exhibit different characteristics than the shadows of spoofed objects (see Figures \ref{fig:real_ped} and \ref{fig:fake_ped}). We use this observation to design an efficient and effective detection mechanism that verifies the presence of objects only when they exhibit the expected 3D shadow effect.

We design, develop and evaluate a system, \systemName{}, which firstly employs ray optics to map the expected \textit{shadow region} of a detected 3D object. Then, it uses a scoring algorithm leveraging exponential decay weight estimation to reduce the importance of measurement artifacts and determine whether the proposed shadow region corresponds to (a) a real shadow or (b) an anomalous shadow. In the latter case, it uses a binary classifier trained on density features extracted from the proposed shadow region, to further classify a shadow region as a \textit{ghost object shadow} or as a \textit{poisoned shadow} (thus verifying the presence of a true object). Our evaluation shows that more than 98\% of the 3D objects in our dataset have meaningful shadows, and that \systemName{}'s shadow region estimation closely captures their true shape. We also show that \systemName{} consistently achieves more than 94\% accuracy in identifying anomalous shadows. \systemName{} can further classify with 96\% accuracy whether the anomalous shadow corresponds to a ghost attack. In addition, we also design a novel, strong, \textit{object invalidation adversary} which follows an optimal strategy to launch an evasion attack that poisons a genuine shadow such that it is misclassified as a ghost shadow, thus invalidating genuine objects. We demonstrate that \systemName{} remains robust. Lastly, \systemName{} achieves real-time detection (2.17x improvement compared to related work~\cite{255240} for processing ghost vehicles) rendering it suitable for deployment both online to provide hints to vehicle passengers, operators or end-to-end AI systems and offline for forensic analysis. Visual examples of \systemName{} are shown online~\cite{sc_website}.

%% file: sections/2_background.tex
\noindent\textbf{LiDAR Sensors.}
\label{subsec:back}
To scan the environment, LiDARs emit a pulse in the invisible near-infrared wavelength (900–1100 nm), which is reflected on incident objects before returning to the receiver. Based on the time of flight, LiDARs calculate the distance between the sensor and the incident object. LiDARs used in AVs (e.g. Velodyne LiDARs) emit a number of light pulses from an array of vertically arranged lasers (16, 32, 64, etc.) that rotate around a center axis to obtain a 360-view of the surroundings of the sensor unit. The sensor translates a return signal to a measurement 3D point consisting of coordinates (x,y,z) and a reflection value (R) corresponding to the return signal's reflectivity or signal strength. 3D point clouds are commonly projected to 2D in a more compact representation called \emph{birds-eye view} or BEV for short.

\vspace{5pt}\noindent\textbf{3D Object Detector Attacks.}
Prior work showed that 3D object detectors based on point-clouds are vulnerable to LiDAR spoofing attacks~\cite{petit2015remote, shin2017illusion, cao2019adversarial, 255240} and point cloud perturbation attacks~\cite{xiang2019generating, zeng2019adversarial, 8803770, yang2019adversarial, wen2019geometry}. Wicker and Kwiatkowska~\cite{wicker2019robustness} further found that 3D object detectors are trained to learn object representations from a ``critical point set'', and subtle changes in the input greatly impact the model's performance. More closely related to our work are LiDAR spoofing attacks. Petit \textit{et al.}~\cite{petit2015remote} first introduced physical attacks to generate noise, fake echos and fake 3D objects by relaying and replaying LiDAR signals. However, they were unable to spoof objects closer than 20m from the LiDAR receiver. Subsequently, Shin \textit{et al.}~\cite{shin2017illusion} managed to inject 10 3D points that are up to 12m in front of the LiDAR receiver. Cao \textit{et al. }~\cite{cao2019adversarial} then demonstrated the capability to spoof up to 100 points and proposed a model-level spoofing methodology that can fool a target AV simulator. Recently Sun \textit{et al.}~\cite{255240} successfully used up to 200 points to launch model-level attacks which leverage 3D points of occluded or distant vehicles to spoof front-near vehicles. Such attacks can have severe repercussions as they can force a vehicle to brake abruptly~\cite{cao2019adversarial}. This might physically injure passengers, halt traffic or induce a crash with the vehicle behind it.

\vspace{5pt}\noindent\textbf{3D Object Detector Defenses.} 
Existing defenses for 3D point cloud object detection focus on defending against point cloud perturbations~\cite{8803770,yang2019adversarial, zhou2019dup}. More related to our work are point injection (or LiDAR spoofing) attacks in AV settings, where suggestions were made to use multi-modal sensor fusion~\cite{ivanov2014attack, yang2018sensor}, view fusion~\cite{255240} and to leverage occlusion patterns~\cite{255240}. Multi-modal sensor fusion makes strong assumptions about the integrity of sensors, Sun \textit{et al.}'s~\cite{255240} SVF approach is promising but not backward compatible, and requires expensive re-training to deal with a noticeable penalty in its classification performance which may prove more dangerous than the attack it tries to address. Sun \textit{et al.}~\cite{255240} also proposed (CARLO) which is similar to our approach as it can be applied orthogonally to the object classifier. However it is significantly slower than \systemName{} for vehicles. More importantly, CARLO is not robust against smaller objects, such as pedestrians and cyclists. Lastly, our work introduces a new semantically meaningful physical invariant, 3D shadows.

%% file: sections/2_1_threat_model.tex
We adopt the threat model from~\cite{cao2019adversarial, 255240} and assume a physical adversary who can spoof LiDAR return signals to fool an AV's 3D object detector model. The adversary can spoof the signals either by placing an attacker device on the roadside or by mounting it on an attack vehicle driving in front of the target vehicle in an adjacent lane~\cite{cao2019adversarial}. The attacker's device can capture the LiDAR signal, and emit a return signal with a delay which controls where in the resulting point cloud the spoofed point will appear. This has been proven as a realistic attack surface~\cite{petit2015remote, shin2017illusion, cao2019adversarial, 255240}. Below we define the adversary's ($\mathcal{A}$) capabilities and goals.

\vspace{5pt}\noindent\textbf{$\mathcal{A}$'s capabilities.} $\mathcal{A}$ enjoys state of the art sensor spoofing capabilities and can inject $\leq200$ points within a horizontal angle of 10$^{\circ}$~\cite{255240} in a 3D scene. $\mathcal{A}$ can launch model-level spoofing attacks able to emulate distant and occluded vehicles~\cite{cao2019adversarial, 255240}. In addition, we consider attacks spoofing smaller objects with $\leq200$ injected points. $\mathcal{A}$ is a white-box adversary with full knowledge of the internals of both the victim model and the detection mechanism.  

\vspace{5pt}\noindent\textbf{Extending the threat model: considering smaller objects.} Prior work introduced a defense (CARLO) against attacks aiming to spoof vehicles \cite{255240}. However CARLO is limited when considering smaller objects, such as pedestrians, cyclists or motorcycles. Our experiments evaluating CARLO's limitations to detect spoofed pedestrian objects are given in Appendix \ref{apx-sec:CARLO_experiment}. Our approach considers vehicle spoofing but also spoofing of smaller objects such as pedestrians and cyclists. The latter is an even easier target for $\mathcal{A}$ but harder to defend; genuine pedestrian and cyclist objects consist of $\sim200$ points on average---by analyzing the full KITTI dataset we found an average number of points of 478, 206 and 174 in the bounding boxes of cars, pedestrians and cyclists respectively. Considering such objects is paramount, as they are commonly encountered, and the safety repercussions can be more severe in the case of an accident. In the KITTI dataset, which contains LiDAR measurements from real-world driving scenarios, more than 30\% of the objects detected on the road are pedestrians and cyclists, with pedestrians being the second most predominant object after cars \cite{Geiger2013IJRR}. Failing to reliably detect (or verify) such objects (e.g. in an invalidation attack), has a higher probability of leading to human injuries and even fatalities \cite{wakabayashi_2018} than accidents involving only vehicles. 

\vspace{5pt}\noindent\textbf{Extending the threat model: object invalidation attacks.}
Using shadows to verify true objects, might incentivize attacks where the adversary’s goal changes from injecting ghost objects to invalidating genuine objects by poisoning their shadow. This could lead to an erroneous safety-critical decision with potentially dire consequences. Our defense mechanism recognizes this. In contrast with related work, it considers for the first time an even stronger adversary which is capable of launching both ghost object injection and object invalidation attacks. To test the robustness of our system against object invalidation attacks, we formulate a novel, strong attack with full knowledge of the detection mechanism, and evaluate the success of this attack against our system.

%% file: sections/3_object_to_shadows.tex
We observe that any 3D object representation in a point cloud is closely followed by a respective region void of measurements. We call this the \textit{3D shadow effect}. Object detectors do not take into account shadow effects and only learn point representations of objects for the detection task. 3D shadow effects occur because LiDAR sensors record measurements (3D points) from return light pulses reflected off an object in a direct line of sight that return within a constrained time period to the receiver of the sensor unit. Thus, anything behind the incident object cannot be reached by the light rays and cannot be measured, resulting in void (shadow) regions. This observation leads us to hypothesize that \textit{the presence and characteristics of shadows can be used to verify genuine 3D objects}. In this section we systematically analyze real 3D driving scenes to verify the presence of shadows in genuine 3D objects, obtain ground truth for such shadow regions and to verify that ghost objects cannot have realistic shadows.

\vspace{5pt}\noindent\textbf{Presence of 3D Shadows.}
To verify the presence of 3D shadows we randomly sampled 120 scenes from the KITTI dataset~\cite{Geiger2013IJRR}. The dataset includes LiDAR measurements (point cloud scenes) from real driving scenarios in Karlsruhe, Germany. The dataset is accompanied by a set of object labels for training 3D object detectors. We used these labels to locate true objects in each scene. We then converted each scene to its birds-eye-view (BEV) representation by projecting each 3D point to a 2D plane. Then, we went through all 120 scenes and (1) manually annotated shadow regions, if present, using the VIA annotation tool \cite{dutta2019vgg}, and (2) assigned shadow regions to objects.  

In the 120 sampled scenes, we found a total of 607 objects (see Table \ref{tab:shadow_breakdown}). All objects are located in the frontal view of the vehicle and include objects both on the road and on sidewalks. Out of the 607 objects, we have identified shadows for 597 or 98.3\% of the objects, the details by object type can be found in Table \ref{tab:shadow_breakdown}. We could not identify shadows for the remaining 10/607 (1.6\%) objects, due to the objects' location in the environment. For example, if one object is directly in front of another but not fully occluding it (e.g., a person is standing in front of a vehicle), the first object cannot be unequivocally assigned a shadow region because of the second object.

\begin{table}[htbp]
\centering
\small
\vspace{-3mm}
\caption{Objects and their shadows in 120 KITTI scenes.}
\label{tab:shadow_breakdown}
\resizebox{0.8\columnwidth}{!}{%
\begin{tabular}{|l|c|c|c|c|}
\hline
                        & \textbf{\begin{tabular}[c]{@{}c@{}}Count\\ In Dataset\end{tabular}} & \textbf{\begin{tabular}[c]{@{}c@{}}\% of total\\ Objects\end{tabular}} & \textbf{\begin{tabular}[c]{@{}c@{}}Labeled\\ Shadows\end{tabular}} & \textbf{\begin{tabular}[c]{@{}c@{}}\% of \\ Object Type\end{tabular}} \\ \hline
\textbf{Car}            & 444                                                                 & 73.1                                                                   & 439               & 98.9                                                                  \\ \hline
\textbf{Pedestrian}     & 45                                                                  & 7.4                                                                    & 41                & 91.1                                                                  \\ \hline
\textbf{Cyclist}        & 17                                                                  & 2.8                                                                    & 17                & 100                                                                   \\ \hline
\textbf{Van}            & 56                                                                  & 9.2                                                                    & 55                & 98.2                                                                  \\ \hline
\textbf{Truck}          & 17                                                                  & 2.8                                                                    & 17                & 100                                                                   \\ \hline
\textbf{Tram}           & 6                                                                   & 1.0                                                                    & 6                 & 100                                                                   \\ \hline
\textbf{Sitting Person} & 1                                                                   & 0.2                                                                    & 1                 & 100                                                                   \\ \hline
\textbf{Miscellaneous}  & 21                                                                  & 3.5                                                                    & 21                & 100                                                                   \\ \hline
\textbf{Total}          & 607                                                                 & N.A.                                                                   & 597               & N.A.                                                                  \\ \hline
\end{tabular}
}
\vspace{-6mm}
\end{table}

\vspace{5pt}\noindent\textit{Conclusion.} By manually labeling shadow regions for objects, we found strong evidence of co-occurrence of objects and shadow regions. This supports our hypothesis that the presence of shadows is a physical invariant that can be potentially used to verify genuine objects in 3D scenes. 

\vspace{5pt}\noindent\textbf{3D Shadows of Genuine vs Ghost Objects.}
LiDARs default operating mode records the \emph{Strongest Return Signal}. Thus, successfully spoofing a signal, is equivalent to transforming a measurement point in the original point cloud to the desired spoofed position~\cite{cao2019adversarial}. In other words, following the physics of the LiDAR, spoofing a point should result in a corresponding point behind the injected point (in the laser ray direction) to be removed from the point cloud. This would result in a void region that might resemble a 3D shadow behind a concentrated attack trace. However, the resolution of LiDAR varies with distance. Ground reflections and objects nearer to the LiDAR have higher density of points. This density decreases as the distance from the sensor increases. Due to the limitations of the attacker's $\mathcal{A}$ capabilities, and the LiDAR's resolution (point density per distance), there exists an effective distance where the attacker would be unable to successfully spoof an object while mimicking a genuine shadow. This presents an opportunity to leverage 3D shadows to develop a robust detection mechanism.

To characterise the LiDAR resolution, we used a scene where the ego-vehicle is in an object-free environment and analysed the ground reflection measurements recorded. A 2m x 2m region of analysis was used to calculate the density of points as a measure of the LiDAR's resolution. We used the objects manually labeled in the KITTI dataset of 7481 scenes, categorised the objects by type, counted the number of points in the object's bounding box and binned them by distance from the ego-vehicle. Table \ref{tab:distance_pts} shows the results of the analysis of average point-cloud density with respect to distance for both ground reflection measurements in clean environment and point measurements in the respective object's bounding box for the whole KITTI dataset. The LiDAR's ground reflection resolution (Row 1 of Table \ref{tab:distance_pts}) decreases sharply as the distance increases. At a distance of 15m-20m, the density is about 200 points in a 2m x 2m region (which corresponds to a horizontal angle of $\leq$ 10$^{\circ}$ at that distance). To have 200 points in their bounding box, Cars would need to be at a distance of 20-25m, whilst Pedestrians and Cyclists would need to be at a distance of 10-15m. Thus, the analysis shows that at a distance of 10m and higher, the resolution of the LiDAR (KITTI data were captured using a Velodyne HDL-64E LiDAR) is insufficient compared to the attacker's capability and the attacker is able to spoof objects with a corresponding shadow region that would be indistinguishable from a real shadow. Therefore, the effective distance of \systemName{} is determined to be 10m which is sufficient to detect the strongest adversary known to date, who tries to spoof a front-near obstacle to cause an unsafe reaction by the AV. Undoubtedly, both LiDAR and adversarial capabilities will evolve. The use of higher-end LiDARs with more laser channels might introduce more attack opportunities, however, the attacker's current capabilities to inject points reliably is limited by hardware. Moreover, higher-end lasers also mean a higher resolution (i.e. a denser point-cloud), which would require the adversary to reliably inject significantly more points to spoof an object that exhibits a realistic shadow. If the adversary's capability matches the LiDAR point density, no defense would be viable.

\begin{longtable}[htbp]{|l|c|c|c|c|c|c|c|c|c|}
\caption{Distance vs Avg. Point Density in Object's Bounding Box and Clean Environment}
\label{tab:distance_pts}\\
\hline
\multirow{2}{*}{}      & \multicolumn{9}{c|}{\textbf{Distance from LiDAR (m) / Avg. Points}} \\ \cline{2-10} 
 &
  \multicolumn{1}{l|}{\textbf{0-5}} &
  \multicolumn{1}{l|}{\textbf{5-10}} &
  \multicolumn{1}{l|}{\textbf{10-15}} &
  \multicolumn{1}{l|}{\textbf{15-20}} &
  \multicolumn{1}{l|}{\textbf{20-25}} &
  \multicolumn{1}{l|}{\textbf{25-30}} &
  \multicolumn{1}{l|}{\textbf{35-40}} &
  \multicolumn{1}{l|}{\textbf{40-45}} &
  \multicolumn{1}{l|}{\textbf{45-50}} \\ \hline
\endfirsthead
\endhead
\textbf{Env (2m x 2m)} & 1295   & 859    & 288   & 196   & 103   & 73    & 42   & 28   & 10  \\ \hline
\textbf{Car (BBox)}    & 4858   & 2040   & 865   & 405   & 208   & 117   & 73   & 51   & 36  \\ \hline
\textbf{Ped (BBox)}    & 1187   & 455    & 207   & 99    & 62    & 42    & 27   & 29   & 16  \\ \hline
\textbf{Cycl (BBox)}   & 1718   & 651    & 263   & 125   & 78    & 53    & 37   & 17   & 12  \\ \hline
\end{longtable}
\vspace{-2mm}

\noindent\textit{Conclusion.} We characterised the LiDAR (Velodyne HDL-64E) scan resolution of ground reflection and objects for the KITTI dataset. The result was used to determine the effective distance which was found to be 10m (conservative), where $\mathcal{A}$ would not be able to reconstruct a legitimate looking object that subverts a shadow detection mechanism. This distance can be further increased with higher resolution LiDARs such as the  Velodyne VLS-128.

%% file: sections/4_ghostbuster.tex
\begin{figure*}[hbtp]
    \vspace{-6mm}
    \centerline{\includegraphics[width=1\textwidth]{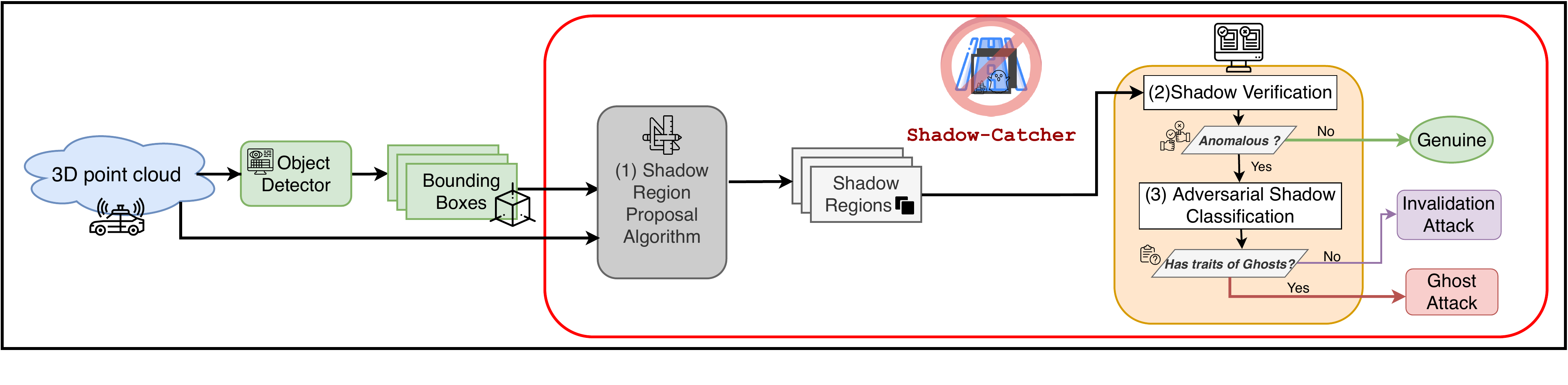}}
    \caption{3D scene perception pipeline with \systemName{} integrated.}
    \label{fig:detection_pipeline2}
    \vspace{-8mm}
\end{figure*}

\vspace{5pt}\noindent\textbf{High Level Architecture.}
\systemName{}'s overall architecture and decision workflow is summarized in Fig.~\ref{fig:detection_pipeline2}. \systemName{} is agnostic to the sensor spoofing methodology and the victim model. It takes as input, the output of a 3D object detector (bounding boxes of detected objects in 3D scene's point cloud) and the original point cloud of the scene, and performs a three-phase analysis to determine whether the detected objects are genuine or ghosts. \systemName{} can further distinguish between ghost objects and genuine objects whose shadow regions are being poisoned. In Phase 1, it employs a \textit{shadow region proposal} algorithm which uses geometrical optics (or ray optics) to generate proposed shadow regions for each of the 3D objects detected by the 3D object detector. By tracing rays from the reference point of the LiDAR unit, \systemName{} can determine the boundaries of shadow regions for 3D objects. However, the shadow region can be imprecise and can include 3D point artifacts which in principle should not be present. To deal with these imprecisions, in Phase-2, \systemName{}'s \textit{genuine shadow verification} component, performs a point-wise analysis in each shadow region to determine whether the region is indicative of a genuine shadow. For this, it uses a novel 3D-point scoring mechanism. If the genuine shadow verification fails, which would mean the system is under attack, \systemName{} uses an \textit{adversarial shadow classification} model to determine whether the shadow region is indicative of a ghost object's shadow (thereby detecting a ghost attack) or a genuine object's shadow (thereby detecting an invalidation attack).
Below we elaborate on \systemName{}'s three main components: (a) shadow region proposal; (b) genuine shadow verification; and (c) adversarial shadow classification.

\vspace{5pt}\noindent\textbf{2D Shadow Region Proposal.}\label{subsec:design_shadow_est} Intuitively, if a scene is converted into its 2D representation, then, using ray optics, we can obtain an area (2D shadow region) behind an object that rays cannot reach since they would have already reflected off the incident surface of the object. To compute the shadow region, we first convert a 3D scene into its 2D birds-eye view (BEV) compact representation. Next, we compute the boundary lines of the shadow region. We first take the coordinates of the bounding box for the detected object from the 3D object detector. Using the coordinates of the corners of the bounding box, we compute the gradients of the lines from the reference point (position of the LiDAR unit) to each of the corners. Let ($x_i$ , $y_i$), $\forall  i = 1,\dots, 4$, be the 4 anchoring corners of a 3D bounding box on the ground. The gradients ($m_i$) of lines connecting the reference point (0,0) to corner coordinates are computed with $m_{i} = \frac{y_i - 0}{x_i - 0}$. The minimum and maximum gradient lines define the \textit{shadow boundary lines} for the shadow region. To simulate the fact that LiDAR has a finite range we define a maximum \textit{shadow length} ($\mathit{l}$). The shadow length ($\mathit{l}$) can be derived from the height of the object ($\mathit{h}$), with respect to the height of the LiDAR unit ($\mathit{H}$) and the furthest distance of the object from LiDAR unit ($\mathit{d_{obj}}$). Using similar triangles, the shadow length ($\mathit{l}$) can be derived as $ l = d_{obj} \times \frac{h}{H-h}$. The shadow boundary lines and the shadow length determine the full shadow region of the object. 

In principle, shadow regions should be completely void of points. Hence, it would suffice to project all the points onto the 2D ground plane before examining the 2D shadow regions. However, 2D shadow regions can only define a 2D area behind the object, which corresponds to the projection of the shadow on the ground. This can result in noisy points contaminating the shadow regions in lieu of taller objects behind the target 3D object.

\vspace{5pt}\noindent\textbf{3D Shadow Region Proposal.} To address the aforementioned limitation of 2D shadow regions we introduce the 3D shadow region estimation. With 3D shadow regions, we examine a volume of space for the presence of points. The base area of the 3D shadow region is the 2D region obtained using the 2D shadow region proposal algorithm. To account for the height of the shadow region, we explore two approaches: (a) simulate LiDAR light rays as in 2D shadow region estimation, but this time analyze the points in the region of space bounded from the ground to the non-occluded incident rays; and (b) use of a uniform height above the ground level to obtain a short volume of space for analysis of points (Illustrations on our website~\cite{sc_website}). The choice of 3D shadow region estimation method and their effects on detection performance are evaluated in Section~\ref{subsec:evaluation_shadow_region_analysis}.

\vspace{5pt}\noindent\textbf{Genuine Shadow Verification.}
\label{sec:detect_algo}
After the shadow regions are identified, \systemName{} performs an analysis inside each region to determine whether the shadow is genuine or not. As mentioned previously, in principle there should be no measurements inside shadow regions, as light rays cannot reach there. However, inaccuracies of the shadow estimation, and noisy artifacts due to physical effects, the placement and shape of objects can result in points being recorded inside genuine shadow regions. Thus, a trivial approach which expects those regions to be completely empty would be frequently flagging real objects as ghosts, resulting in high error rates.


To mitigate this, we propose a method which reduces the significance of noisy measurements inside shadow regions and assigns the shadow region an anomaly score. Our genuine shadow verification method classifies a shadow as genuine if its anomaly score is below a threshold. It first assigns a weight to each point inside the shadow region. Intuitively, points due to noise are assigned a lower weight, but points found in non expected regions (i.e. along the center-line or close to the start-line) are assigned higher weights. Specifically, we use two exponential decay equations (Eq. \eqref{eq:weight_start} and \eqref{eq:weight_mid}) on two axis of analysis to assign weights to the points, where $x_{start}$, $x_{end}$ , $x_{mid}$ and $x_{bound}$ are the distances of the point from the start-line, end-line, center-line and closest boundary line of the shadow region and $\alpha$ is a parameter that tunes the rate of exponential decay. The aggregate anomaly score of the shadow region is computed using Eq. \eqref{eq:score}, where $w_{min}$ is the minimum weight a point can obtain in any axis of analysis (i.e. point at boundary line) and $T$ is the total number of points in shadow.

\begin{equation}\label{eq:weight_start}
    \small
    w_{start} = \text{exp} \left( \frac{\ln(0.5)}{\alpha} \times \frac{x_{start}}{(x_{start} +x_{end})} \right)
\end{equation}

\begin{equation}\label{eq:weight_mid}
    \small
    w_{mid} = \text{exp} \left( \frac{\ln(0.5)}{\alpha} \times \frac{x_{mid}}{(x_{mid} +x_{bound})} \right)
\end{equation}

\begin{equation}\label{eq:weight_min}
    \small
    w_{min} = \text{exp} \left( \frac{\ln(0.5)}{\alpha} \right)
\end{equation}

\begin{equation}\label{eq:score}
    \small
    score = \frac{ \sum_{i=1}^{T}(w_{start,i} \times w_{mid,i}) - (T \times w_{min}^2)}{T \times (1-w_{min}^2)}
\end{equation}

The anomaly score threshold is set empirically. An extensive analysis was performed and the Receiver Operating Characteristic (ROC) curve was used to determine the threshold that produces an acceptable True Positive and False Positive Rate (see Section~\ref{subsec:evaluation_shadow_region_analysis}). An object is verified as genuine by \systemName{} if its shadow region gets a lower score than the anomaly threshold, otherwise the shadow is flagged as anomalous. At this point \systemName{} can already detect that the system is under a LiDAR poisoning attack. However, we take this a step further and also try to determine the type of attack against the system.

\vspace{5pt}\noindent\textbf{Adversarial Shadow Classification.}
A high shadow anomaly score indicates either a ghost attack or an object invalidation attack. \systemName{} distinguishes between the two. We observe that during ghost attacks, the shadow regions of ghost objects exhibit a high density of points while points are sparse in the shadow regions of true objects during invalidation attacks. Therefore, we expect the distribution of points within the shadow regions of ghost vs invalidated objects to be distinguishable. Leveraging these observations we use clustering to extract density features from shadow regions, which we then use to train a binary adversarial shadow classifier. \systemName{} uses this classifier to determine whether an anomalous shadow is the result of a ghost attack or an invalidation attack.

\vspace{5pt}\noindent\textit{Feature Extraction.}
To characterize the density of the measurements in a shadow region, we cluster points that are in spatial proximity. We use ``Density-Based Spatial Clustering of Applications with Noise" (DBSCAN)~\cite{ester1996density} for this purpose as it is able to identify points that are clustered in arbitrary shapes and does not require to pre-specify the number of clusters. This suits our use-case well, as point clusters in 3D point clouds are irregular and the number of clusters in a region is not known a priori. 

Clustering points in shadow regions with DBSCAN, allows us to extract the number of clusters found by controlling the density of clusters. Intuitively we would expect the shadow regions of ghost objects to exhibit multiple clusters with regular and similar shapes. On the other hand, during an object invalidation attack, we would expect the shadow region to be mainly void  with points injected by the attacker eliciting a high aggregated score near the region of high weighting as modeled by the exponential decay equations in the axis of analysis. Thus, a distinguishable characteristic of shadows for an invalidation attack would be a small number of or no clusters detected (Example in \cite{sc_website}). From DBSCAN, we then derive the following features to characterize the shadows of objects: (a) \textit{number of clusters} in the shadow region obtained from DBSCAN; (b) \textit{average density of points in clusters} obtained by taking the total number of points in clusters and averaging out by the number of clusters.

\vspace{5pt}\noindent\textit{Attack Classification Model.} The shadow characteristic features are then used as input to a binary classification model to distinguish between shadows of ghost objects and shadows of genuine objects under an invalidation attack. Note that the attacker can elicit a high anomaly score by opportunistically injecting a single point at the shadow location of highest weighting. Whilst this triggers the anomaly detection, it fails to elicit a ghost attack classification. To defeat the mechanism, an attacker would have to effectively emulate shadows representative of ghost attack shadows, which requires both injecting points at regions of high weighting and creating multiple clusters with sufficient density of points (i.e. to emulate the shadow features of ghost shadows). We define the \textit{object invalidation attack} in the following sub-section.

\vspace{5pt}\noindent\textbf{Object Invalidation Attack.}
\label{subsec:object_invalidataion_attack}
\systemName{}'s use of shadows, can incentivize a new class of object invalidation attacks targeting genuine objects' shadows. We formulate this as an evasion attack (test-time) on the adversarial shadow classification model. We consider a strong adversary who has state-of-the-art LiDAR spoofing capabilities and knowledge of the classifier's decision boundary and feature representation (i.e. shadow characteristics features). Their goal is to introduce points in the shadow region of a genuine object to change the shadow's characteristics and cause the classification model to misclassify a genuine shadow as a ghost object shadow, effectively invalidating the real object. 

We can evaluate the robustness of the classification according to the capability of the adversary. In our case, we define the attacker's capability as the total number of points that can be injected in a target shadow region in a single point cloud scene. We refer to this as the adversary's ``point budget". We then define the invalidation adversary's budget $B_{\mathcal{A}}$ as:
\begin{equation}
\small
     n_0 + n_p = N_{c} \times \rho_{c}, \ \ \ \text{s.t.} \ n_p \leq B_{\mathcal{A}}
\label{eqBudget}
\end{equation}
where $n_0$ is the original number of points in the shadow region, $n_p$ is the number of injected malicious points, $N_c$ is the number of clusters after injection, and $\rho_c$ is the average cluster density after injection. 

Intuitively, the invalidation adversary's optimal strategy against \systemName{} can be defined as follows: Given a set of features for a genuine shadow, inject the minimum number of points, $n_p$, to deceive the classifier by modifying the combination of cluster density and number of clusters, subjected to a point budget $B_{\mathcal{A}}$ and the configuration parameters of DBSCAN used by \systemName{}. This optimal attack strategy can be formalized as:
\begin{equation}
\small
\min n_p , \ \  \text{s.t.} \ \exists \ N_c \in \mathcal{Z}^+  | \  F( (n_0+n_p)  / N_c, N_c) = 1   
\label{eqEvasion}
\end{equation} where $F(\cdot,\cdot)$ is the output of the classifier, which is one if an attack is identified as a ghost and zero otherwise. As the complexity of the optimization problem in Eq.~(\ref{eqEvasion}) is reduced, $n_p \in \mathcal{Z}^+$ is a scalar and the classifier has only 2 features, the problem can be solved using simple techniques such as the bisection method. We evaluate the robustness of \systemName{} to such an adversary in Section \ref{sec:results_evaluation}.

%% file: sections/5_results_evaluation.tex
We evaluate \systemName{}'s effectiveness and efficiency in detecting ghost and invalidation attacks. We also evaluate its accuracy in estimating shadow regions, but due to space limitations we present that analysis in Appendix~\ref{apx-sec:2d_shadow}.

\vspace{3pt}\noindent\textbf{Ghost Object Injection for \systemName{} Evaluation.} \systemName{} is agnostic to the adversarial strategy and is applied on the output of 3D object detectors. Therefore, it suffices to evaluate its response on the products of object spoofing attacks: bounding boxes of detected spoofed objects and resulting point cloud. Our attacks follow $\mathcal{A}$'s capabilities and generate large (cars) and small (pedestrians, cyclists) spoofed objects in front-near locations of 5-8m in-front of the victim vehicle. To create a ghost object of a particular type (e.g. pedestrian), we first extract the point clouds of genuine objects from real-world point clouds (from KITTI~\cite{Geiger2013IJRR}) and prune them to $\mathcal{A}$'s capabilities, the maximum physical capabilities demonstrated in the related work (200 points within a spoofing angle of 10$^\circ$). The resulting attack trace is then added into the target scene's point cloud. We then remove any points behind the attack trace (to obey the LiDAR single return signal measurement mode) effectively recreating the result of a real-world spoofing attack. For each object type, its attack trace is injected in 200 random scenes from the KITTI dataset, resulting in an \textit{Adversarial Dataset} containing a total of 600 scenes (200 scenes $\times$ 3 objects injected).

\begin{figure}[h]
\vspace{-4mm}
    \centerline{\includegraphics[width=0.9\textwidth]{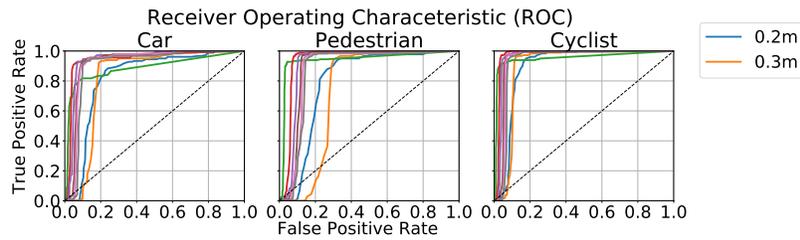}}
    \caption{ROC curves for detection of injected objects.}
    \label{fig:roc1}
    \vspace{-5mm}
\end{figure}

\vspace{5pt}\noindent\textbf{Anomalous Shadow Detection.} \label{subsec:evaluation_shadow_region_analysis}
We used the Adversarial Dataset to evaluate \systemName{}'s ability to correctly detect ghost objects' shadows as anomalous and real objects' shadows as non-anomalous. Ground truth labels of scenes together with ghost object label were used for the bounding box generation. We evaluated Shadow-Catcher's scoring method using 2D shadow Regions (BEV) and 3D Shadow Regions (Ray Height and Uniform Height with different height values ranging from 0.1m to 0.6m above ground level). Our main results are summarized in Figure~\ref{fig:roc1}. A more detailed report on our evaluation is presented in \cite{sc_website} due to space limitations.

\noindent\textit{2D vs 3D Shadow Regions.} We found that 3D shadow regions with uniform height above the ground level outperform the other shadow regions of interest. Noisy points (stray reflections or overhanging objects such as tree branches and sign posts) are reduced when we only consider a small volume above ground which only captures the LiDAR scan reflections off the ground when there are no objects. This results in a more accurate scoring of the shadow region to detect anomalous shadows that have points in sub-regions of high weighting.

\vspace{5pt}\noindent\textit{Height Sensitivity Analysis and Anomaly Detection Threshold.} For shadow regions with uniform height above ground, we performed a sensitivity analysis to find the optimal height for detecting ghost objects. The height that yields the best AUC-ROC is 0.2m above the ground with a score of 0.94, 0.95 and 0.96 for detection of injected Car, Pedestrian and Cyclist respectively. Shadow regions with uniform height of 0.2m also result in the maximum F1 Score and Accuracy for the detection of injected objects. Finally, we parametarized the detection threshold and found that an anomaly score threshold of 0.2 provide the best trade-off of TPR vs FPR for all injected objects.

\vspace{5pt}\noindent\textit{Utility.} Overall, we found that 3D shadows, with a uniform height of 0.2m above the ground and an anomaly score threshold of 0.2 provide the best overall trade-off of TPR vs FPR for all injected objects. In particular, with the above configuration, we obtain an overall accuracy of 0.94, TPR of 0.94 and FPR of 0.069 for anomalous shadow regions due to Ghost Attacks. This is comparable to CARLO's performance~\cite{255240}, which reports a reduced attack success rate of 5.5\% for Car objects. CARLO was not evaluated on smaller objects~\cite{255240}, however our experiments suggest that it will severely underperform(see Appendix~\ref{apx-sec:CARLO_experiment}).

We also took a closer look into the sources of errors (illustration of examples in \cite{sc_website}). False negatives (ghost objects not detected by \systemName{}) in the dataset can typically be attributed to injected objects being implanted in regions that are already void of points due to incomplete LiDAR measurements. Note that this is due to the LiDAR's failure to take measurements of the ground level. As LiDAR technology gets better, our approach's accuracy will also improve. False positives (real objects flagged as potential ghosts by \systemName{}) are due to their shadows having points from other larger objects behind them. Although, this may (very rarely) happen, the safety repercussions are less important as the second object (right behind the first) will be correctly validated.

\vspace{5pt}\noindent\textbf{Attack Classification.} \label{subsec:evaluation_gb_robustness}
Next, we evaluate \systemName{}'s ability to distinguish between ghost attacks and object invalidation attacks. We first show how the features extracted from shadow regions by \systemName{}'s differ between shadows of ghost and genuine objects, and that they can be used to train an effective binary classifier. Then we evaluate \systemName{}'s classification robustness against a novel, strong invalidation adversary with LiDAR spoofing capabilities and full knowledge of the classification method aiming to misclassify a genuine shadow as a ghost shadow.

\noindent\textit{Feature Characteristics of Shadows.}
\label{subsubsec:evaluation_rq3_shadow_characteristics}
We used genuine object shadows for comparison with ghost object shadows for two reasons. (a) \systemName{}'s anomaly detection might (but very rarely) incorrectly mark a genuine object's shadow as anomalous. Being able to distinguish between the two, acts as a second line of validation which can correct \systemName{}'s mistake in the previous phase. This can lead to better utility. (b) An invalidation attack adversary targets genuine object shadows. Thus the distinction between genuine and ghost shadows can serve as a baseline for detecting against invalidation attacks (in the next subsection we use this baseline to design a strong invalidation adversary).

We use the Adversarial Dataset for evaluation. \systemName{} was used to generate object shadows of uniform height (0.2m) and then we compute the \textit{number of 3D point clusters} in each shadow region and \textit{density of the clusters} using DBSCAN ($\epsilon=$0.2, min\_pts=6). The shadows are labeled (ghost vs genuine) and split into a training set and a test set (80:20) to train six different binary classifiers and evaluate their performance on the test set. As we are most concerned with the TPR and FPR of classifiers for best utility, we use AUC-ROC as the decision criteria to choose the model. 

We found that the prevalence rate of feature combination of 0 clusters and 0 cluster density for genuine shadows is 86.2\% (1899/2202) and for ghost shadows is 4.4\% (24/543). We further observe that 91.7\% (2020/2202) of the genuine object shadows have $<5$ clusters and 95.4\% (2101/2202) have $<10$ clusters. Of those genuine shadows with clusters, 91.6\% (2019/2202) have average clusters density of $<10$ points and 98.6\% (2172/2202) have average cluster densities of $<20$ points. These genuine shadow regions are opportunities for an adversary to perform an invalidation attack. A least effort adversary will target shadows which will likely be incorrectly marked as anomalous or force triggering anomaly detection with a single 3D point injected in sub-regions of high importance.

We trained six classifiers and compare their performance (Table~\ref{tab:classifier_performance}). We chose these classifiers as they tend to do well for low dimensionality data. We found that a SVM Model with with polynomial degree=2 provided the best AUC-ROC performance in distinguishing between ghost and genuine shadows.
\begin{table}[htbp]
\small
\vspace{-6mm}
\centering
\caption{Performance Metrics for Shadow Classifiers}
\label{tab:classifier_performance}
\begin{tabular}{|c|c|c|c|}
\hline
                             & \textbf{Accuracy} & \textbf{F1-Score} & \textbf{AUC-ROC}  \\ \hline
\textbf{Logistic Regression} & 0.962             & 0.909             & 0.953            \\ \hline
\textbf{Random Forest}       & 0.969             & 0.921             & 0.938            \\ \hline
\textbf{SVM- Linear}         & 0.960             & 0.906             & 0.978            \\ \hline
\textbf{SVM-Poly(deg=2)}     & 0.965             & 0.918             & \textbf{0.981}   \\ \hline
\textbf{SVM-Poly(deg=3)}     & 0.967             & 0.923             & 0.971            \\ \hline
\textbf{SVM-RBF}             & 0.974             & 0.938             & 0.967            \\ \hline
\end{tabular}
\vspace{-4mm}
\end{table}


\vspace{5pt}\noindent\textbf{Robustness Against Evasion Attacks on Shadow Classification Model.}

\vspace{5pt}\noindent We further evaluate the robustness of the adversarial shadow classifier against the invalidation adversary defined in Section \ref{subsec:object_invalidataion_attack}, setting the DBSCAN parameters as before ($\epsilon=$0.2, min\_pts =6). 
To visualize the maximum cluster-density combination the attacker can introduce given a budget $B_{\mathcal{A}}$ (Eq. \ref{eqBudget}), we use the \textit{Maximum Operating Curve} (MOC), which shows the set of valid $(\rho_c, N_c)$ combinations on the feature space that can be reached for a given $B_{\mathcal{A}}$ (we use 20, 40, 60, 100 and 200 points). In our previous experiments we found that 86.2\% of the genuine shadows have cluster-density combination on the feature space of value 0 for both $N_c$ and $\rho_c = (n_0+n_p)  / N_c$. Thus, for solving the problem in Eq.~(\ref{eqEvasion}), we start exploring this cluster-density combination on the feature space.

\begin{figure}[h]
    \vspace{-5mm}
    \centerline{\includegraphics[scale=0.25]{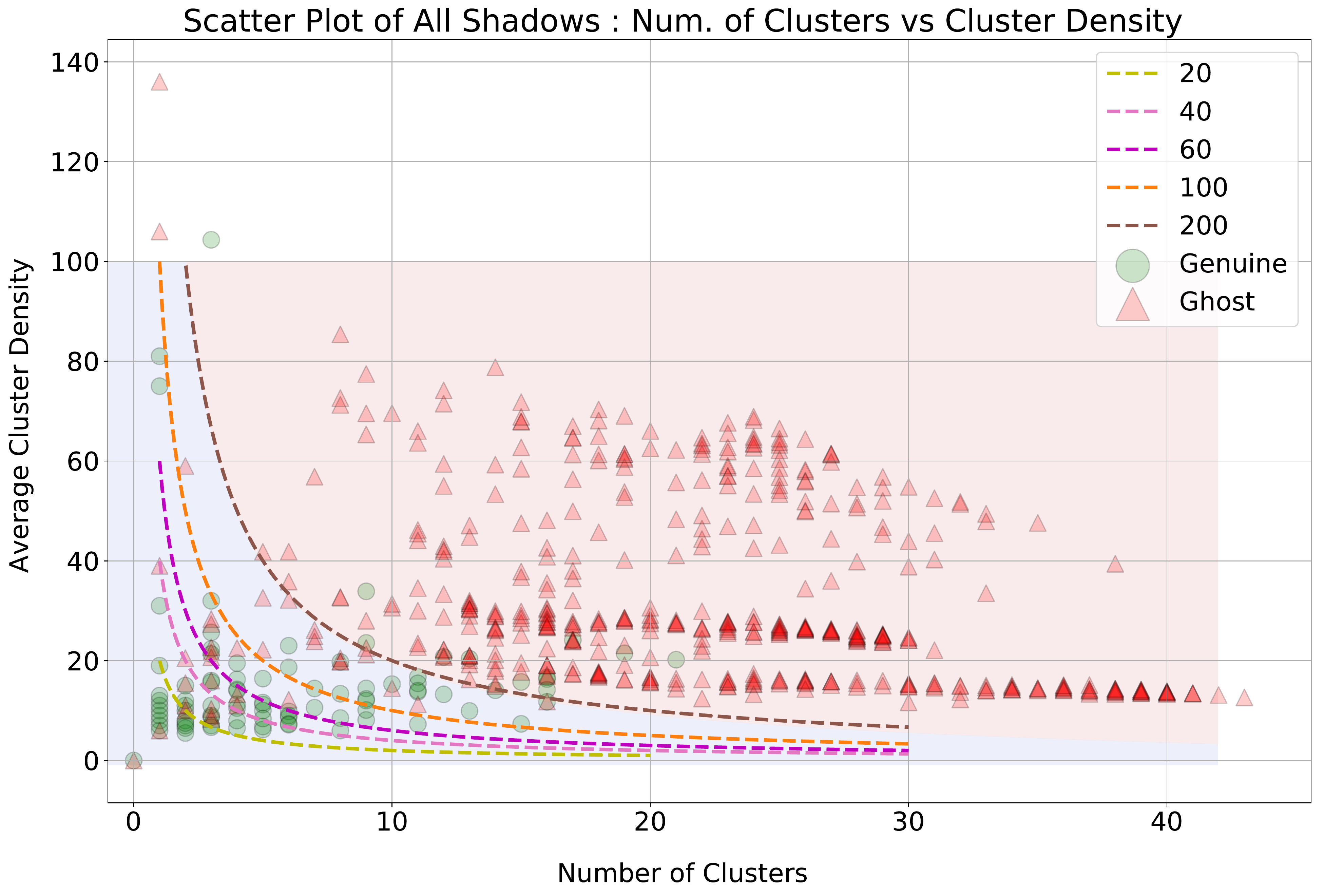}}
    \caption{Scatter plot of shadow features and decision regions from SVM Classifier with a polynomial kernel with degree = 2. Dashed lines are operating curve of adversary according to their budget from (0,0).}
    \label{fig:shad_scatter_poly2}
    \vspace{-5mm}
\end{figure}

This is a very strong adversary because we assume that: (a) the adversary can predict the optimal cluster-density combinations; (b) the adversary can identify where the 3D points should be introduced in the environment to achieve that combination; (c) it is feasible to introduce those measurements. 

Fig. \ref{fig:shad_scatter_poly2} shows all shadow features and the decision regions from the boundary of the non-linear SVM classifier (poly deg=2). A point in the red region is the feature combination where the classifier model will label the shadow as a ghost attack and blue region an invalidation attack. The dashed curves represent MOCs\footnote{The MOC contains discrete values, given the set of valid combinations $(\rho_c, N_c)$. For illustration purposes, we plot the MOC as a continuous contour. } for different budgets from the origin (0,0), which is the  predominant feature combination (86.2\%) for all genuine shadows in our dataset. We observe that for MOCs of up to 200 points, the curves lie within the decision region for non-ghost shadows. Thus, an attacker needs to inject 200 or more points to evade the non-linear SVM model which shows that our model is robust against an adversary that can reliably spoof up to 200 points. With technological improvements, adversarial capabilities can potentially improve beyond 200 points, but there are also ample opportunities to improve Shadow-Catcher accordingly. When higher resolution LiDARs are used, \systemName{} robustness improves, as long as adversarial spoofing capabilities do not match the LiDARs’ resolution.

\vspace{5pt}\noindent\textbf{Runtime Efficiency.}\label{subsec:evaluation_efficiency}
To evaluate \systemName{}'s runtime efficiency, we use the same Adversarial Dataset. We measure \systemName{}'s end-to-end analysis time for each identified object (genuine and ghost), starting from the time \systemName{} receives the 3D objects bounding box coordinates until \systemName{} labels the object. We also measure the execution time for each component of \systemName{}. \systemName{} is configured to use 3D shadow generation with a uniform height of 0.2m  above ground,  an  anomaly score threshold of 0.2, DBSCAN for feature extraction with $\epsilon=$0.2, min\_pts=6, and our pre-trained SVM binary classifier with a polynomial kernel of degree=2. \systemName{}'s prototype implementation is written in Python with 1200 lines of code. We measure the execution time on a machine equipped with an Intel Core i7 Six Core Processor i7-7800X (3.5GHz) and 32GB RAM.

\begin{table}[h]
\vspace{-6mm}
\centering
\small
\caption{\systemName{}'s object processing times (in ms)}
\label{tab:gb_time}
\resizebox{0.8\columnwidth}{!}{%
\begin{tabular}{|l|c|c|c|c|}
\hline
                    & \textbf{\begin{tabular}[c]{@{}c@{}}Shadow\\ Generation\end{tabular}} & \textbf{\begin{tabular}[c]{@{}c@{}}Shadow\\ Scoring\end{tabular}} & \textbf{\begin{tabular}[c]{@{}c@{}}Shadow\\ Verification\end{tabular}} & \textbf{\begin{tabular}[c]{@{}c@{}}Total\\ Time\end{tabular}} \\ \hline
\textbf{Car}        & 0.4$\pm$0.3                                                                      & 4$\pm$10                                                                     & N.A.                                                                          & 4.4$\pm$10.3                                                             \\ \hline
\textbf{Pedestrian} & 0.3$\pm$0.1                                                                      & 6$\pm$8                                                                      & N.A.                                                                          & 6.4$\pm$8.1                                                               \\ \hline
\textbf{Cyclist}    & 0.3$\pm$0.1                                                                      & 3$\pm$4                                                                      & N.A.                                                                          & 3.3$\pm$4.1                                                               \\ \hline
\textbf{Car (ghost)}  & 0.4 0.1                                                                      & 10$\pm$6                                                                     & 10.06$\pm$20.05                                                                   & 20.46$\pm$26.15                                                           \\ \hline
\textbf{Ped. (ghost)} & 0.3 0.1                                                                      & 7$\pm$5                                                                      & 10.06$\pm$17.05                                                                   & 17.36$\pm$22.15                                                           \\ \hline
\textbf{Cyc. (ghost)} & 0.3$\pm$0.1                                                                      & 7$\pm$6                                                                      & 12.06$\pm$24.05                                                                   & 19.36$\pm$24.15                                                           \\ \hline
\end{tabular}
}
\vspace{-4mm}
\end{table}

Table \ref{tab:gb_time} summarizes our results. On average, \systemName{} processes objects in a scene in 0.003s--0.021s. This is only a small fraction of the time a 3D object detector takes to analyze a scene---Point-GNN has an average inference time of 0.6s~\cite{geiger}. Genuine objects are processed much faster than adversarial objects, which is important as this corresponds to most frequently encountered cases. The longer duration taken to process adversarial object shadows is mainly due to the feature extraction step, which is triggered when a shadow is deemed anomalous by the shadow scoring mechanism, requiring 10.7ms on average. The variation observed in the total execution time, comes from the different object sizes and the different point densities in their shadows. Notably, \systemName{} can process a spoofed car in 46.6ms (worst case), which compared to prior work (100 ms on average)~\cite{255240} constitutes at least a 2.17x speedup. Moreover, \systemName{}'s is implemented in Python and thus, can be readily improved even further if more efficient languages are used (e.g. C).

%% file: sections/8_conclusion.tex
In this work we introduced 3D shadows, a new semantically meaningful physical invariant for verifying the presence of objects in a 3D scene. Then we introduced a set of new techniques embodied in an end-to-end system (\systemName{}) which leverage shadows to tackle spoofing attacks against 3D object detectors. Our evaluation shows that \systemName{} achieves 94\% and 96\% average accuracy in identifying anomalous shadows and classifying them as either ghost or invalidation attacks. We further design a strong, novel invalidation adversary aiming to evade classification and found that \systemName{} remains robust up to 200 points. \systemName{} can analyze objects in real time (0.003s–0.021s on average, a  2.17x speedup compared to the state of the art).

%% file: appendix/appendix.tex
\section{Limitation of prior art}\label{apx-sec:CARLO_experiment}
Recently, Sun \textit{et al.} proposed CARLO~\cite{255240}, a system for detecting model-level LiDAR spoofing attacks. CARLO consists of two components. The first, Laser Penetration Detection (LPD), serves as a quick anomaly detector to filter fake and valid objects. Objects for which LPD is not confident in its decision are sent for further analysis to a second component, the Free Space Detection (FSD), which is computationally more expensive. LPD's design intuition is that points in the frustum correlate with occlusion patterns, and hence, uses the ratio of the number of points behind the object's bounding box over the total the number of points in the frustum of an object; objects with high ratio are classified as suspicious or definitely fake. This approach uses points in the bounding box (as part of the frustum), and for smaller objects, the ratio is small and heavily influenced by noisy LiDAR measurements. Moreover, the approach does not take into account the location and characteristics of points in the region behind the bounding box, and could be susceptible to false positives from noise artifacts. FSD's detection is based on the intuition that genuine vehicles have high density of points and hence, low free space in the bounding boxes as most of the space in the bounding box should be occluded by points in front. However, for smaller objects, this approach might be ineffective as the original space in the bounding box is small and mostly occupied by the points. Hence there are limited regions for analysis of free space. We implemented CARLO and evaluated its effectiveness to distinguish genuine from spoofed pedestrian objects.

\vspace{5pt}\noindent\textbf{LPD evaluated on Pedestrians.} To evaluate the LPD ratios of genuine and spoofed pedestrians, we collected the LPD ratios of genuine pedestrian objects in the KITTI dataset as well 200 spoofed front-near pedestrians (6m in front of ego-vehicle). Fig. \ref{fig:lpd_distribution_ped} shows the distribution of LPD ratios of genuine and spoofed pedestrians. We observe that there is an overlap of the two distributions from 0.5 to 0.8, which presents opportunities for attackers to invoke FSD. Additionally, as the LPD ratio's denominator accounts for all the points in the frustum, and for small objects the number of points in frustum is small, there is a possibility of an attacker to inject points (within the total adversary $\mathcal{A}$ budget) in the frustum to lower the ratio to trigger FSD.

\begin{figure}
    \centering
    \begin{minipage}{0.45\textwidth}
        \centering
        \includegraphics[width=0.9\textwidth]{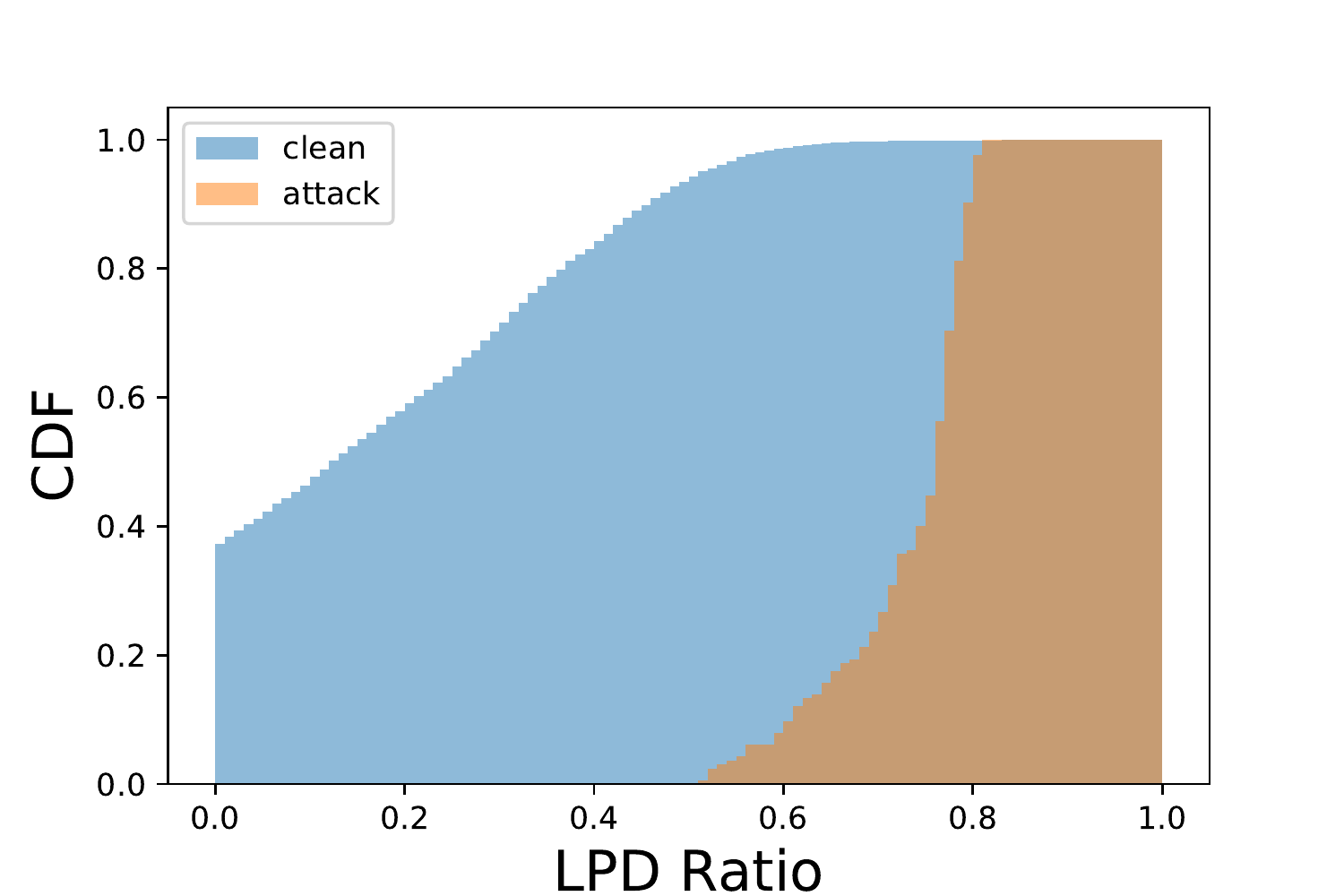} 
        \caption{LPD ratio distribution of Genuine and Spoofed Pedestrian Objects.}
         \label{fig:lpd_distribution_ped}
    \end{minipage}\hfill
    \begin{minipage}{0.45\textwidth}
        \centering
        \includegraphics[width=0.9\textwidth]{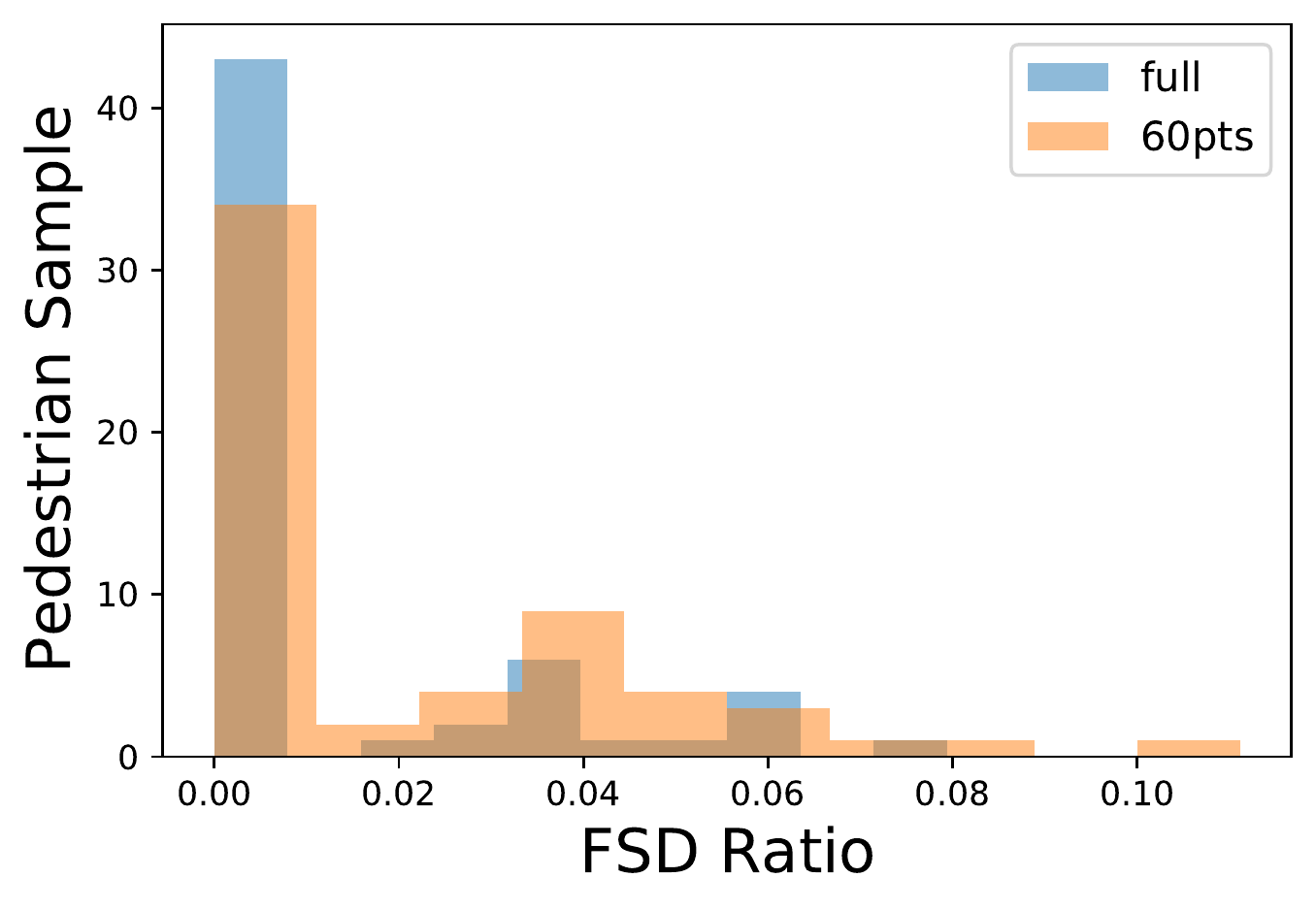} 
        \caption{FSD ratio distribution of Pedestrian Objects with Full and Down-sampled Point Cloud.}
        \label{fig:fsd_distribution_ped}
    \end{minipage}
    \vspace{-4mm}
\end{figure}

\ignore{
\begin{figure}[htbp]
    \vspace{-4mm}
    \centerline{\includegraphics[scale=0.5]{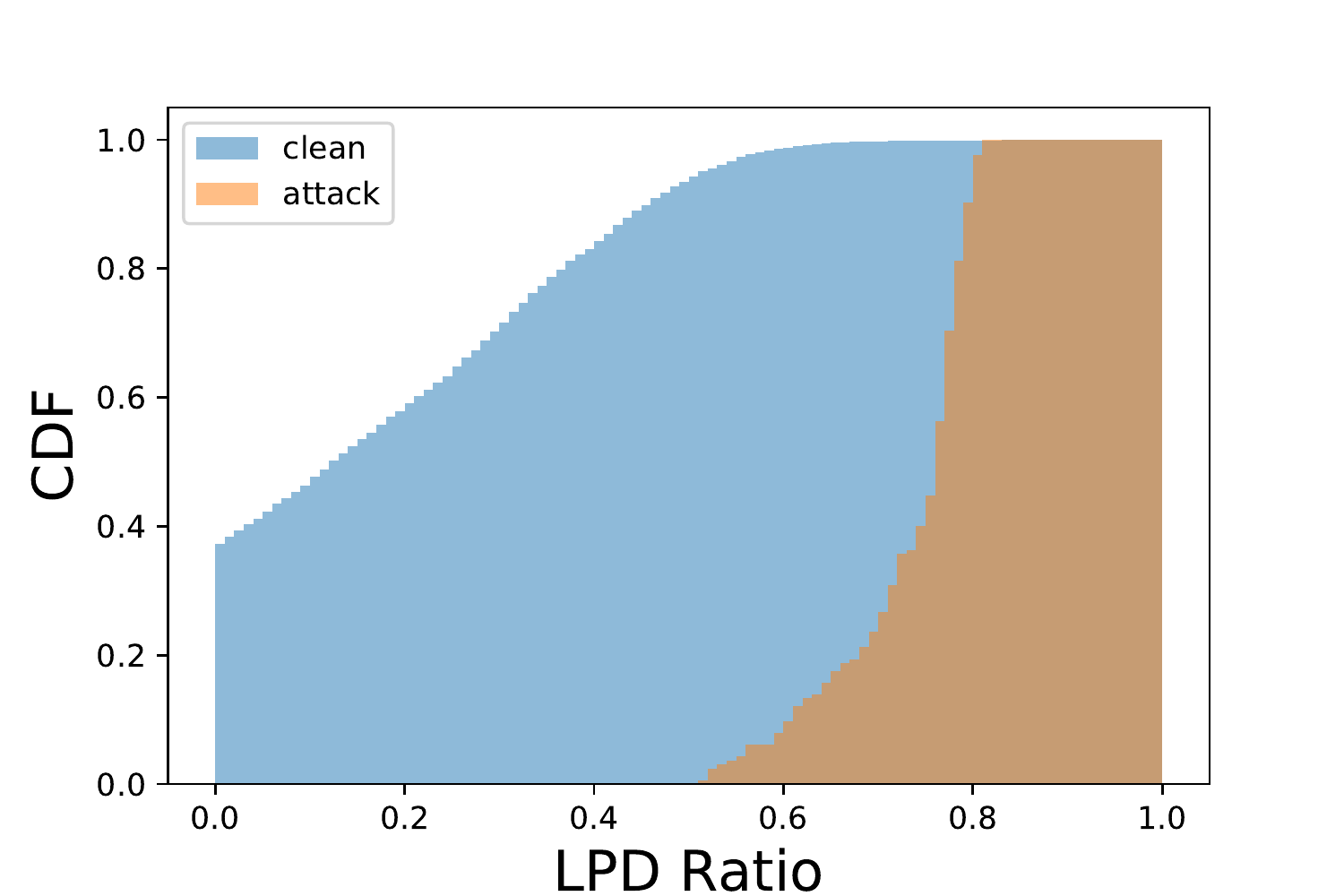}}
    \caption{LPD ratio distribution of Genuine and Spoofed Pedestrian Objects.}
    \label{fig:lpd_distribution_ped}
    \vspace{-4mm}
\end{figure}

\begin{figure}[htbp]
    \centerline{\includegraphics[scale=0.5]{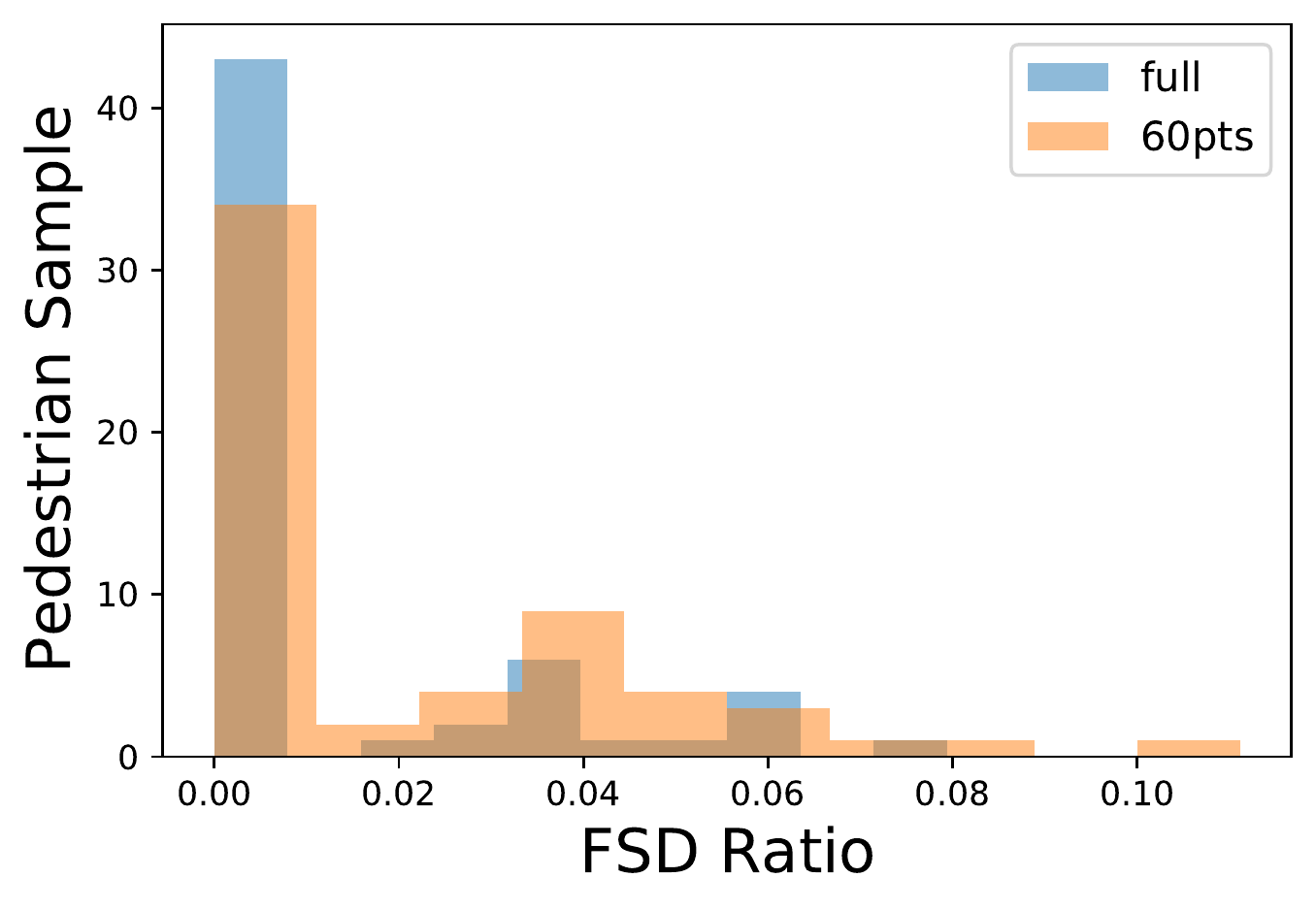}}
    \caption{FSD ratio distribution of Pedestrian Objects with Full and Down-sampled Point Cloud.}
    \label{fig:fsd_distribution_ped}
    \vspace{-4mm}
\end{figure}
}

\vspace{5pt}\noindent\textbf{FSD evaluated on Pedestrians.} We randomly sampled 60 genuine pedestrian objects from KITTI and injected their point cloud 6m in front of the ego-vehicle to spoof a front-near obstacle. Using the same 60 pedestrian objects, these objects' point clouds were also down-sampled to the size of 60 points (below adversary $\mathcal{A}$'s budegt of 200 points) and were similarly injected (Point-GNN detected all down-sampled traces as pedestrians). We then used the implementation of Free Space Detection (FSD) in CARLO to evaluate the FSD ratio of spoofed objects with the full-sized and down-sampled point clouds. Fig. \ref{fig:fsd_distribution_ped} shows that the distribution of FSD ratio overlaps for pedestrians objects of full and down-sampled point clouds, with the majority of them having an FSD ratio of 0. This shows that the approach of FSD will not result in a separable distribution to effectively distinguish small spoofed pedestrians from genuine pedestrians. FSD expects ghost objects to result in very high FSD ratios which as we showed does not happen for small objects.

\section{2D Shadow Region Estimation.}\label{apx-sec:2d_shadow}
\ignore{Algorithm~\ref{alg:2d} shows how to compute the 4 corners of a 2D shadow region in the ground plane.} We analyze \systemName{}'s accuracy of 2D shadow region generation by comparing it with the 597 manually labeled shadows (see Section~\ref{sec:obj_shadow}). We evaluate the 2D region generation separately since 3D regions build on top of it. The significance of 2D vs 3D region estimations in the detection performance is evaluated separately in Subsection~\ref{subsec:evaluation_shadow_region_analysis}. To quantify how closely \systemName{} can match the objects' observed shadows, we measure their Intersection over the Union (IoU) and perform a Procrustes shape analysis. An IoU value of 1 means that the two regions are perfectly matched and 0 means the two regions are disjoint. Procrustes provides us with two metrics: (a) similarity of the shapes; and (b) scale differences of the shapes \cite{gower1975generalized, ross2004procrustes, procrustes_analysis_matlab}. For similarity, values close to 1 mean that the shapes are identical. For scale, a value of 1 means that the size of the shapes are identical and anything less than 1 means the ground-truth shadow shape is smaller, and larger than 1 is the opposite.

\begin{table}[htbp]
\vspace{-5mm}
\begin{center}
\caption{Aggregated correspondence metrics of all objects}
\label{tab:shadow_metrics}
\begin{tabular}{|l|c|c|c|c}
\cline{1-4}
                            & \textbf{IoU} & \textbf{Similarity} & \textbf{Scale} &  \\ \cline{1-4}
\textbf{Mean}               & 0.728        & 0.713               & 1.286          &  \\ \cline{1-4}
\textbf{Median}             & 0.760        & 0.969               & 0.970          &  \\ \cline{1-4}
\textbf{Standard Deviation} & 0.152        & 0.376               & 2.08           &  \\ \cline{1-4}
\end{tabular}
\end{center}
\vspace{-6mm}
\end{table}

Table~\ref{tab:shadow_metrics} summarizes our results across all object types. Detailed results are deferred to the project website\cite{sc_website}. From the median values of the corresponding metrics, it can be observed that, for more than half the objects, the computed shadow matches closely with the ground-truth shadow---IoU, Similarity and Scale values are well above 0.5 which indicates a good prediction (object detection bounding box accuracy is commonly evaluated at IoU $\geq0.5$ ~\cite{redmon2016you,everingham2010pascal}). We do observe some variation in the results which can be attributed to measurement inaccuracies and human-errors in the labeling process, and to over-estimation of shadow areas (Illustration provided on the project's website \cite{sc_website}). \systemName{} uses bounding boxes which are larger than the actual objects and this results in larger shadow regions. However, \systemName{}'s exponential decay approach to weighting the significance of 3D points in shadows (see Section~\ref{sec:system}) compensates for this. This is verified with \systemName{}'s overall accuracy in detecting genuine shadows, ghost and invalidation attacks (see Section~\ref{sec:results_evaluation}). 